# A Novel Energy Resolved X-Ray Semiconductor Detector

Tengfei Yan [1], Chunlei Yang [2], Xiaodong Cui [1*]

[1.] Physic Department, University of Hong Kong, Hong Kong, China

[2.] Shen Zhen Institutes of Advanced Technology, Chinese Academy of Sciences, Shenzhen, China

**The hyperspectral X-ray imaging has been long sought in various fields from material analysis to medical diagnosis. Here we propose a new semiconductor detector structure to realize energy-resolved imaging at potentially low cost. The working principle is based on the strong energy-dependent absorption of X-ray in solids. Namely, depending on the energy, X-ray photons experience dramatically different attenuation. An array or matrix of semiconductor cells is to map the X-ray intensity along its trajectory. The X-ray spectrum could be extracted from a Laplace like transform or even a supervised machine learning. We demonstrated an energy-resolved X-ray detection with a regular silicon camera.**

Since the discovery in 1895, X-ray technology has brought a widespan revolution in fundamental sciences from astronomy, materials science, biomedical sciences, and in daily technologies from security inspection to medical diagnostic imaging. X-ray is a family of electromagnetic waves with photon energies of $10^2$-$10^6$ eV, equivalent to $10^0$-$10^{-3}$ nm in wavelength. For energy of X-ray photons is orders of magnitude higher than that of the out-shell electron transitions, X-ray experiences weak light-matter scattering and consequently shows a long penetration depth compared with infrared and visible lights. This makes X-ray a unique position in undestructive structure/material analysis and medical imaging.

As the X-ray sources generally span in a wide range in spectrum, the energy resolved detection has been demanded and realized by various techniques including energy disperse via crystal grating, metal filters, multi-scintillators, photon counting combined with scintillators and photon counting with fast drift detectors or complementary metal-oxide semiconductor (CMOS) detectors.[1-4] Despite the impressive achievements, these established techniques suffer from several drawbacks. Thus, the implantation of energy resolved X-ray detectors is very limited. As a matter of fact, the X-ray imaging in medical diagnosis is still overwhelmingly in monochrome form.

In this report we propose a new X-ray detector structure to realize energy resolved detection and imaging. The working principle is based on the pronounced energy-dependent attenuation of X-ray in solids. Illustrated with an example of silicon as shown in Figure 1a, its X-ray attenuation coefficient is a strong function of photon energy. As the photoelectric effect which dominates the X-ray absorption for photons of less than 200KeV shows a remarkable energy dependence: its cross section $\sigma \propto \frac{Z^{4\sim5}}{E^3}$ where $Z$ is the atomic number and $E$ is X-ray photon energy.[5] This energy dependent absorption leads to distinct extinction patterns along the X-ray photon trajectory at different energies. Starting with a conceptual detector consisting of arrays of pixels/cells as shown in Figure 1b and c, one could approximate the signal at individual cells generated by absorption and/or scattering X-ray photon by

$$I_i(z) \sim \int_0^\infty \rho(\omega)(e^{-\alpha(\omega)Z} - e^{-\alpha(\omega)[Z+\Delta Z]})f(\omega)d\omega = \sum I(\omega)(e^{-\alpha(\omega)Z} - e^{-\alpha(\omega)[Z+\Delta Z]})f(\omega) \quad (1)$$

Where $I_i(z)$ is the signal at the *i*-th cell with the distance of $Z$ from the surface; $\Delta Z$ is the cell width; $\rho(\omega)$ and $\alpha(\omega)$ denote the spectrum profile intensity/weight and the attenuation coefficient at photon energy $\omega$;

$I(\omega)$ the X-ray spectral intensity at the energy window $(\omega - \delta/2, \omega + \delta/2)$, and $f(\omega)$ describes the cell signal yield at energy $\omega$.

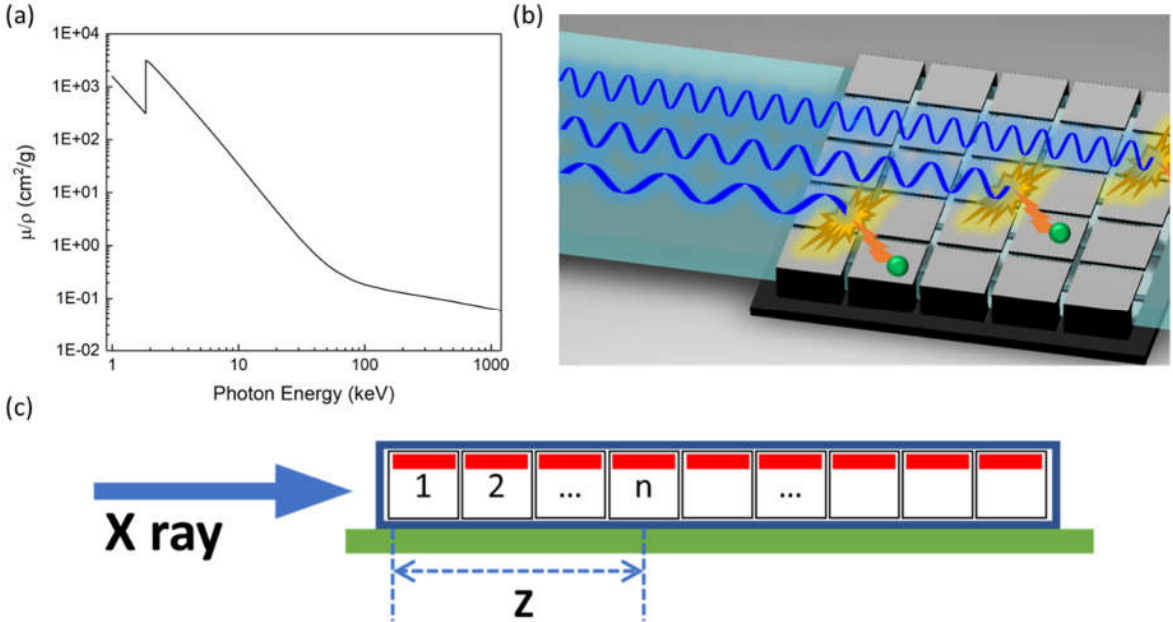

Figure 1. (a) Mass attenuation coefficient μ/ρ as a function of photon energy of silicon.[6] (b) Schematic diagram of a conceptual device. The X-ray photons sheds in parallel along the array/matrix of detector cells. The energy dependent attenuation coefficient leads to energy-dependent decay patterns along the X-ray photon trajectory. (c) Front view of the conceptual device, consisted of an array of detector cells (from cell 1 to cell n). Z denotes the distance of the cell from the front surface of the detector. X-ray photons travel along the array and vanish by energy transfer along the cells.

Depending on the cell size and energy resolution requirements, one can set the total cell number which equivalently set the upper limit of X-ray energy resolution. Namely, the more cells, the higher energy resolution. For example, if one picks the total cell number of *n*, the energy resolution then is set at the X-ray energy range to *n* windows. So we get a linear equation set

$$\begin{pmatrix} I(1) \\ I(2) \\ I(3) \\ \vdots \\ I(n) \end{pmatrix} = \begin{bmatrix} (e^{-\alpha(\omega_1)Z_1} - e^{-\alpha(\omega_1)[Z_1+\Delta Z]})f(\omega_1)^1 & (e^{-\alpha(\omega_2)Z_1} - e^{-\alpha(\omega_2)[Z_1+\Delta Z]})f(\omega_2)^1 & \cdots & (e^{-\alpha(\omega_n)Z_1} - e^{-\alpha(\omega_n)[Z_1+\Delta Z]})f(\omega_n)^1 \\ (e^{-\alpha(\omega_1)Z_2} - e^{-\alpha(\omega_1)[Z_2+\Delta Z]})f(\omega_1)^2 & (e^{-\alpha(\omega_2)Z_2} - e^{-\alpha(\omega_2)[Z_2+\Delta Z]})f(\omega_2)^2 & \cdots & (e^{-\alpha(\omega_n)Z_2} - e^{-\alpha(\omega_n)[Z_2+\Delta Z]})f(\omega_n)^2 \\ \vdots & \vdots & & \vdots \\ (e^{-\alpha(\omega_1)Z_n} - e^{-\alpha(\omega_1)[Z_n+\Delta Z]})f(\omega_1)^n & (e^{-\alpha(\omega_2)Z_n} - e^{-\alpha(\omega_2)[Z_n+\Delta Z]})f(\omega_2)^n & \cdots & (e^{-\alpha(\omega_n)Z_n} - e^{-\alpha(\omega_n)[Z_n+\Delta Z]})f(\omega_n)^n \end{bmatrix} \begin{pmatrix} I(\omega_1) \\ I(\omega_2) \\ I(\omega_3) \\ \vdots \\ I(\omega_n) \end{pmatrix}$$

Where *I(1), I(2),…* denote the signal at the *1st, 2nd,… n*-th detector cells, respectively; and $f(\omega_i)^j$ is the signal yield at the energy around $\omega_i$ of the *j*-th cell. If the secondary photons, electron-hole pairs migration, Compton scattering, *etc.* are neglected, the $f(\omega_i)^j$ is roughly a constant across the cell array. Theoretically the X-ray spectrum *I(ω)* at specific energy range centering around *ω* could be analytically extracted by

$$\begin{pmatrix} I(\omega_1) \\ I(\omega_2) \\ I(\omega_3) \\ \vdots \\ I(\omega_n) \end{pmatrix} = \begin{bmatrix} (e^{-\alpha(\omega_1)Z_1} - e^{-\alpha(\omega_1)[Z_1+\Delta Z]})f(\omega_1)^1 & (e^{-\alpha(\omega_2)Z_1} - e^{-\alpha(\omega_2)[Z_1+\Delta Z]})f(\omega_2)^1 & \cdots & (e^{-\alpha(\omega_n)Z_1} - e^{-\alpha(\omega_n)[Z_1+\Delta Z]})f(\omega_n)^1 \\ (e^{-\alpha(\omega_1)Z_2} - e^{-\alpha(\omega_1)[Z_2+\Delta Z]})f(\omega_1)^2 & (e^{-\alpha(\omega_2)Z_2} - e^{-\alpha(\omega_2)[Z_2+\Delta Z]})f(\omega_2)^2 & \cdots & (e^{-\alpha(\omega_n)Z_2} - e^{-\alpha(\omega_n)[Z_2+\Delta Z]})f(\omega_n)^2 \\ \vdots & \vdots & & \vdots \\ (e^{-\alpha(\omega_1)Z_n} - e^{-\alpha(\omega_1)[Z_n+\Delta Z]})f(\omega_1)^n & (e^{-\alpha(\omega_2)Z_n} - e^{-\alpha(\omega_2)[Z_n+\Delta Z]})f(\omega_2)^n & \cdots & (e^{-\alpha(\omega_n)Z_n} - e^{-\alpha(\omega_n)[Z_n+\Delta Z]})f(\omega_n)^n \end{bmatrix}^{-1} \begin{pmatrix} I(1) \\ I(2) \\ \vdots \\ I(n) \end{pmatrix}$$

Thus, a hyperspectral X-ray image could be acquired with the above proposed device. This method eliminates the challenging detector performance requirements for time response and absorption efficiency which usually are contradicting in device design. With the decreased size for individual cells, the demanding for high purity of bulky semiconductor crystals is much relaxed compared to photon counting based X ray detectors.

To validate this structure, we test with a commercial CMOS camera (Sony alpha 6300, Silicon CMOS size 23.5 x 15.6 mm, 24M pixels, shown in figure 2a) under exposure of a compact X-ray tube with tungsten target and .5mm Be window. The working current is set at 0.8 mA. Figure 2b shows the cross profile (the intensity vs. pixel position) along the X-ray propagation direction under X-ray exposure at various energy. The profile clearly shows distinct exponential decay patterns for X-ray photons with different energies. As the X-ray source carries a broad band spectrum, the profile could be analyzed as a superposition of X-ray absorption curves along the CMOS detector at a continuous energy range. The X-ray spectrum shifts to higher energy range at higher X-ray tube bias, and the attenuation across silicon dramatically decreases as the X-ray photon energy increases. So the X-ray displays distinct decay pattern at different X-ray tube bias.

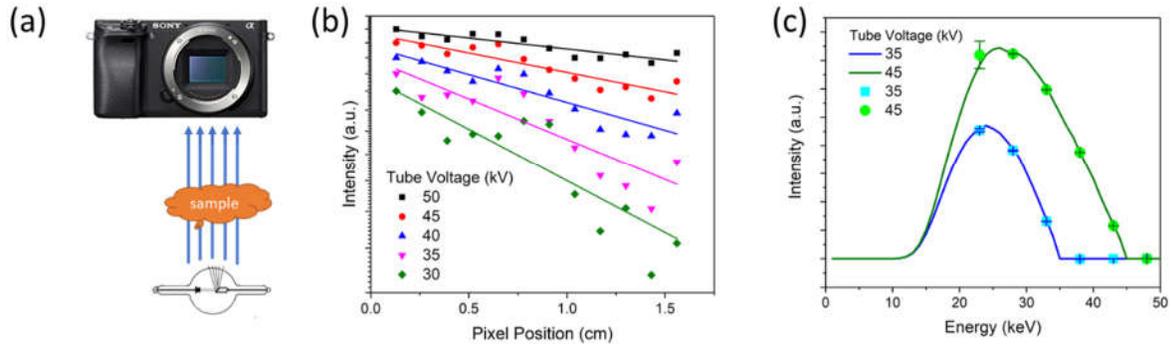

Figure 2. (a) Schematic geometry of the demo experiment. (b) Camera response as a function of the distance from the X-ray entrance surface at various X-ray tube voltages. The solid lines present the fitting assuming single exponential decay, indicating different penetration length at different energy X-ray. (v) Converted spectra from the camera response calculated with transform matrix achieved by linear regression method (Scikit-learn package). The solid lines shows the emission spectrum from the X-ray tube calculated with spektr 3.0 package.[7]

The transform matrix which converts the signal as a function of pixel position to the energy resolved signals could be analytically calculated if the signal yield $f(\omega_i)^j$ at specific pixel is identified. Given that the signal generation is affected by various mechanisms besides the dominating photoelectric effect, for example, Compton scattering, electron-positron generation, secondary photon scattering, etc., the $f(\omega_i)^j$ may not be constants across all the pixels. Ideally the detector $f(\omega_i)^j$ could be measured under the exposure to monochromatic X-ray beams scanning across the energy range. An alternative and easier approach towards the transform matrix is through the regression via supervised machine learning. The CMOS signal profiles under the exposure of the known spectrum were used as the training data set. The regression methods of linear least square model, least absolute shrinkage and selection operator (LASSO), Least-angle regression (LARS) were exploited and all gave the converging transform matrix. Figure 2c shows a representative transform of a spectrum with 6 discrete energy windows ranging from 25KeV to 45keV after 12 training sets. Cross validation gives an error smaller than 7% across the whole spectrum range.

Figure 3 demonstrates an energy-dependent imaging of a dummy sample of copper wires on top of aluminum sheet with the CMOS camera. Figure 3b shows the X-ray image under a X-ray tube voltage of 40KV, where the binned intensity along the X-ray propogation gives a integrated response of the broadban X-ray. The images at various energy windows can be extracted simutaneously with the transform matrix as demonstrated above. Figure 3c shows the image of the low energy compoents ($23\pm5.5$keV) where both copper and aluminum parts efficently attenute the low energy X-ray, resulting in blur constrast; While figure 3d, the image of the high energy component ($34\pm5.5$ keV) shows significantly improved contrast, resulting from remarkable diffference in energy attenutation between copper and aluminum. Figure 3e illustrates the CMOS response for the transmitted X-ray from aluminnum (Point A) and copper (Point B). The contrast of the images at different energies is well expected from the distinct X-ray attenuation in different materials. This approach towards energy dependent X-ray imaging would be an economical way for applications in material characterization and medical imaging.

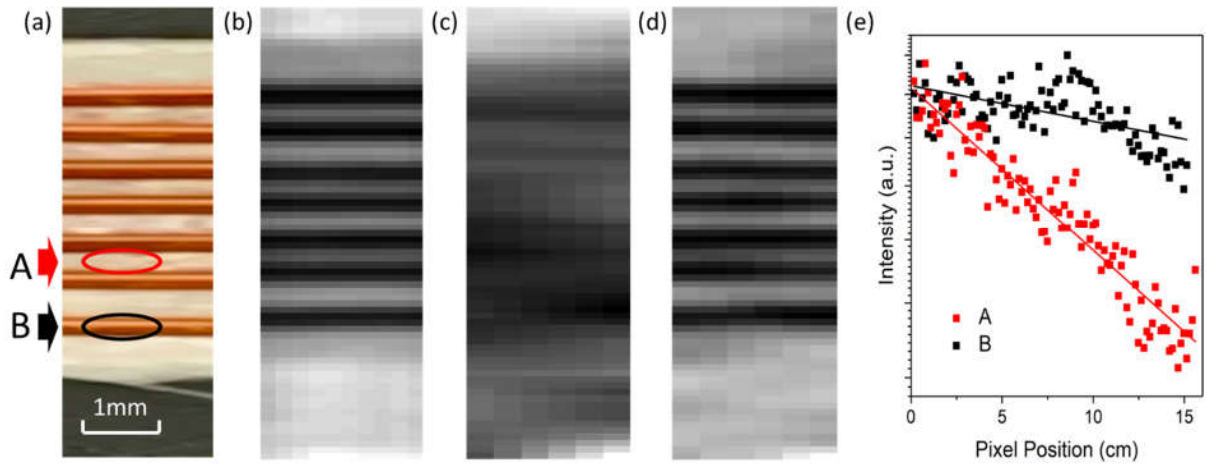

Figure 3. (a) Photo of the dummy sample of copper wires on top of an aluminum flake. (b) X-ray images taken with demo camera under exposure of an X-ray tube with bias voltage of 40 kV. (c) and (d) are images of the low (23±5.5keV) and high energy (34±5.5keV) X-ray components, extracted from the energy integrated image (b) with the transform matrix, respectively. (e) camera response as a function of pixel position along X-ray penetration direction for transmitted X-ray at spot A and B labeled by black and red arrows in figure 3(a). Solid lines are exponential decay function fittings for eye guidance.

In summary we demonstrate a new pixel-based X-ray detector structure to realize efficient energy resolution with high efficiency. The basic mechanism utilizes the energy dependent attenuation coefficient in X-ray energy range and the spatial-energy transform. The energy resolved imaging could be realized with a matrix of cell structure which could be in a form of 2D panel stacking in either perpendicular or parallel direction. The proposed structure offers a new route to realize low cost X-ray spectrometer and energy resolved X-ray imaging.

**Acknowledgement:** The project was supported by Croucher Foundation.